\documentclass[letterpaper, 10 pt, conference]{ieeeconf}  

\IEEEoverridecommandlockouts                              
\usepackage{color}
\usepackage{amssymb}  
\usepackage{booktabs}
\usepackage{tabularx}
\newlength\subtabdist
\usepackage[utf8]{inputenc}
\usepackage{graphicx}
\usepackage{xcolor}
\usepackage{subfig}

\usepackage{amsmath}
\usepackage{MnSymbol}
\usepackage{comment}

\usepackage{tikz}
\usetikzlibrary{arrows,intersections}

\usepackage[utf8]{inputenc}
\newtheorem{theorem}{Theorem}
\newtheorem{problem}{Problem}
\newtheorem{definition}{Definition}
\newtheorem{properties}{Property}
\newtheorem{proposition}{Proposition}
\newtheorem{remark}{Remark}
\newtheorem{exm}{Example}
\usepackage{pgfplots} 
\pgfplotsset{compat=newest} 
\pgfplotsset{plot coordinates/math parser=false} 
\newlength\figureheight 
\newlength\figurewidth 
\usepackage[strict=true,style=english]{csquotes}

\newcommand{\argmin}{\mathop{\mathrm{argmin}}} 

\newcommand{\argmax}{\mathop{\mathrm{argmax}}} 
\newcommand{\sign}{\mathop{\mathrm{sign}}}

\usetikzlibrary{shapes,snakes}

\title{\LARGE \bf{Specifying User Preferences using Weighted Signal Temporal Logic
}}

\author{Noushin Mehdipour$^{1}$, Cristian-Ioan Vasile$^{2}$ and Calin Belta$^{1}$
\thanks{This work was partially supported at Boston University by the NSF under grant IIS-1723995. ${}^1$Noushin Mehdipour (noushinm@bu.edu) and Calin Belta (cbelta@bu.edu) are with the Division of Systems Engineering at Boston University, Boston, MA, USA, and ${}^2$Cristian-Ioan Vasile (cvasile@lehigh.edu) is with the Department of Mechanical Engineering and Mechanics, Lehigh University, Bethlehem, PA, USA.}%
}
\begin{document}
\maketitle
\thispagestyle{empty}
\pagestyle{empty}

\begin{abstract}
We extend Signal Temporal Logic (STL) to enable the specification of importance and priorities. The extension, called Weighted STL (wSTL), has the same qualitative (Boolean) semantics as STL, but additionally defines weights associated with Boolean and temporal operators that modulate its quantitative semantics (robustness). We show that the robustness of wSTL can be defined as weighted generalizations of all known compatible robustness functionals (i.e., robustness scores that are recursively defined over formulae) that can take into account the weights in wSTL formulae. 
We utilize this weighted robustness to distinguish signals with respect to a desired wSTL formula that has sub-formulae with different importance or priorities and time preferences, and demonstrate its usefulness in problems with conflicting tasks where satisfaction of all tasks cannot be achieved. We also employ wSTL robustness in an optimization framework to synthesize controllers that maximize satisfaction of a specification with user specified preferences. 
\end{abstract}
\section{Introduction}
\label{intro}Temporal logics, such as Linear Temporal Logic (LTL) and Computation Tree logic (CTL) \cite{baierBook} are formal specification languages that enable expressing temporal and Boolean properties of system executions.
Recently, temporal logics have been used to formalize specifications for complex monitoring and control problems in cyber-physical systems. A variety of tools has been developed for analysis and control of many systems from such specifications~\cite{belta,tabuada2009verification,alur1996hybrid}. 

Signal Temporal Logic (STL)~\cite{stl} specifies signal characteristics over time. Its quantitative semantics, known as robustness, provides a measure of satisfaction or violation of the desired temporal specification, with larger robustness indicating more satisfaction.
The quantitative semantics enables formulating STL satisfaction as an optimization problem with robustness as the objective function. This problem has been solved using heuristics, mixed-integer programming or gradient methods~\cite{CDC},~\cite{milp,raman},~\cite{pant2017smooth,belta2019formal}.%

Multiple functionals have been proposed to capture the STL quantitative robustness.
The traditional robustness introduced in~\cite{donze} uses $\min$ and $\max$ functions over temporal and logical formulae, resulting in an extreme, \emph{sound}, non-convex and non-smooth robustness function. 
For linear systems with linear costs and formulae, traditional robustness optimization approaches commonly encoded Boolean and temporal operators as linear constraints over continuous and integer variables~\cite{milp,raman}. However, the resulting Mixed Integer Linear Programs (MILPs) scaled poorly with the size and horizon of the specifications (i.e., they require a large number of  
integer variables). 
Later works employed smooth approximations for $\max$ and $\min$ to achieve a differentiable robustness and use scalable gradient-based optimization methods applicable to general nonlinear systems. However, the \textit{soundness} property was lost due to the approximation errors~\cite{pant2017smooth}. 

Several works have tackled the issue of defining sound robustness functionals with regularity properties (i.e., continuity and smoothness)~\cite{discrete,iman19,varnai,gilpin2020smooth}. 
In~\cite{ACC}, the limitations of traditional robustness (induced by the $\min$ and $\max$ functions) in optimization were categorized as {\em locality} and {\em masking}. {\em Locality} means that robustness depends only on the value of signal at a single time instant, while {\em masking} indicates that the satisfaction of parts of the formulae different from the most ``extreme" part does not contribute to the robustness.
~\cite{ACC,CDC19} employed additive and multiplicative smoothing and eliminated the locality and masking effects to enhance optimization. Later works \cite{varnai,gilpin2020smooth} defined parametric approximations for $\max$ and $\min$ that enabled adjustment of the locality and masking to a desired level. A similar issue was studied in LTL specifications, where a counting method was used to distinguish between small and large satisfactions (or violations) of a LTL formula \cite{tabuada2015robust}. 

All these works have focused on the run-time performance of the planning or verification 
with temporal logic specifications. However, little attention has been devoted to the problem of capturing user preferences in satisfying temporal logic properties with timing constraints. In LTL, specifying the preferences of multiple temporal properties was addressed in \emph{minimum-violation} \cite{rrt} or \emph{maximum realizability} \cite{dimitrova2018maximum} problems, i.e., if multiple specifications are not realizable for a system, it is preferable to synthesize a minimally violating or maximally realizing system. These problems were formulated by assigning priority-based positive numerical weights (weight functions) to the LTL formulae \cite{dimitrova2018maximum} or corresponding deterministic transition systems \cite{rrt}. 
The idea of using priority functions was also studied in \cite{hoxha2018mining} in order to prioritize optimization of specific parameters in a mining problem with parametric temporal logic properties. Time Window Temporal Logic (TWTL) proposed in \cite{twtl} enabled specifying preferences on the deadlines through temporal (deadline) relaxations and formulation of time delays.

However, for STL specifications, the problem of capturing user preferences, i.e., \emph{importance} or \emph{priorities} of different specifications or the timing of satisfaction is not well understood. The contributions of this paper are: (1) we extend STL to \emph{Weighted Signal Temporal Logic (wSTL)} to formally capture importance and priorities of tasks or timing of satisfaction via weights; (2) we show that the extended quantitative semantics can be defined as a weighted generalization of a recursively defined STL robustness functional, 
(3) we propose adapted evaluation and control frameworks that use wSTL to reason about a system behavior with incompatible (infeasible) tasks or with performance preferences. 
\section{Preliminaries}
\label{preliminaries}
Let $f:\mathbb{R}^n \rightarrow \mathbb{R}$ be a real function. We define $[f]_ + = {\small \begin{cases} f & f > 0\\ 0 & \text{otherwise}\end{cases}}$ and $[f]_- = - [ - f]_+$, where $f =[f]_+ + [f]_-$.
The sign function is denoted by $\sign:\mathbb{R} \to \{-1, 0, 1\}$.

\subsection{Signal Temporal Logic (STL)}
STL was introduced in~\cite{stl} to monitor temporal properties of real-valued signals.
Consider a discrete- or continuous-time domain $\tau \subseteq \mathbb{R}^+$.
A \textit{signal} $S$ is a function $S:\tau\rightarrow\mathbb{R}^n$ that maps each time point $t \in \tau$ to an $n$-dimensional vector of real values $S(t)$. 
We denote $I=[a,b]:= \{t \in \tau \mid a\leq t \leq b\}$ 
and $t+I$ as the interval $[t+a, t+b]$.
The STL syntax is defined and interpreted over $S$ as follows:\vspace{-3pt}
\begin{equation}
\label{eq:syntax}
\varphi:=\top \mid \mu \mid \neg\varphi \mid \varphi_1 \land \varphi_2 \mid \mathbf{F}_{I} \varphi \mid \mathbf{G}_{I} \varphi, \vspace{-3pt}
\end{equation}
where $\varphi$, $\varphi_1$, $\varphi_2$ are STL formulae, $\top$ is logical \emph{True}, $\mu:=(l(S(t))\geq 0)$ is a \emph{predicate} where $l: \mathbb{R}^n \to \mathbb{R}$ is a Lipschitz continuous function defined over the values of $S$, and $\lnot$ and $\land$ are the Boolean \emph{negation} and \emph{conjunction} operators. The Boolean constant $\bot$ (\textit{False}) and the other Boolean operators (e.g., \emph{disjunction} operator $\lor$) can be defined from $\top$, $\lnot$, and $\land$ in the usual way. Temporal operator \emph{eventually} $\mathbf{F}_{I}\varphi$ is satisfied if \enquote{$\varphi$ is True at some time in $I$};
while \emph{always} $\mathbf{G}_{I}\varphi$ means \enquote{$\varphi$ is True at all times in $I$}. For example, formula $\varphi= \mathbf{G}_{[0,7]}\mathbf{F}_{[0,3]}(S>0)$
specifies that for all times between 0 and 7, within the next 3 time units, signal $S$ becomes positive. STL qualitative semantics determines whether $S$ satisfies $\varphi$ at time $t$ ($S\models^t \varphi$) or violates it ($S\nmodels^t \varphi$). 
Its quantitative semantics, or \emph{robustness}, measures \emph{how much} a signal satisfies or violates a specification.
\begin{definition}[Traditional Robustness \cite{donze}]
\label{def:trad-robustness}
Given a specification $\varphi$ and a signal $S$, the traditional robustness
$\rho(\varphi,S,t)$ at time $t$ is recursively defined as follows \cite{donze}:\vspace{-3pt}
\begin{equation}
\label{eq:trad-robustness}
\begin{aligned}
\rho ( {\mu, S, t} ) &: =  l(S(t)),\\
\rho \left( {\neg \varphi, S, t} \right) &: =  - \rho (\varphi, S, t),\\
\rho \left( {\varphi_1 \land \varphi_2 ,S,t} \right) &:= \min \left( \rho(\varphi_1, S, t), \rho(\varphi_2, S, t) \right),\\
\rho \left( {\varphi_1 \lor \varphi_2 ,S,t} \right) &:= \max \left( \rho(\varphi_1, S, t), \rho(\varphi_2, S, t) \right),\\
\rho \left( {{\mathbf{G}_{I}} \varphi, S, t} \right) &: = \mathop{\inf} \limits_{t' \in{t + I}} {\rho(\varphi, S, t')},\\
\rho \left( {{\mathbf{F}_{I}} \varphi, S, t} \right) &: = \mathop{\sup} \limits_{t' \in{t + I}} {\rho(\varphi, S, t')}.
\end{aligned}
\vspace{-3pt}
\end{equation}
\end{definition}
\begin{theorem}[Soundness \cite{donze}]
The traditional robustness is sound, i.e., $\rho \left( {\varphi ,S,t} \right) > 0$ implies $S\models^t \varphi$, and $\rho \left( {\varphi ,S,t} \right) < 0$ implies $S\nmodels^t \varphi$.
\end{theorem}
\vspace{-2pt}
We call STL sub-formulae $\varphi_i$ connected by a conjunction operator \emph{obligatory}, i.e., all $\varphi_i$s must be satisfied for $\varphi={\bigwedge_{i}\varphi_i}$ to be satisfied. We also call STL sub-formulae $\varphi_i$ connected by the a disjunction operator \emph{alternative}, i.e., $\varphi={\bigvee_i\varphi_i}$ is satisfied if either one of $\varphi_i$ is satisfied.

\subsection{Weighted Arithmetic and Geometric Means}\label{sec:wmean}
Weighted arithmetic and geometric means of a finite set 
$\mathbf{x}=\{ x_1,x_2,...,x_m\}$ with corresponding non-negative weights $\mathbf{w}=\{w_1,w_2,...,w_m\}$ with $\sum\limits_{i = 1}^m w_i = 1$ are given by:\vspace{-6pt}
\begin{equation*}\small
\begin{array}{c}
\bar{\mathbf{x}}_{\text {Arithmetic}}= \sum\limits_{i = 1}^m {w_i x_i},\quad 
\bar{\mathbf{x}}_{\text {Geometric}} = \prod\limits_{i = 1}^m { x_i^{w_i}}
= \exp \left(\sum\limits_{i = 1}^m {w_i \ln x_i} \right)
 \end{array}  
 \vspace{-3pt}
\end{equation*}
\subsection{Smooth Approximations}
The $\max$ and $\min$ functions can be approximated by:\vspace{-2pt}
\begin{equation}
\label{eq:smooth}
\begin{array}{l}
\widetilde{\min}\{x_1,\ldots,x_m\}:=-\frac{1}{\beta}\ln (\sum_{i=1}^m e^{- \beta x_i}),\\
\widetilde{\max}\{x_1,\ldots,x_m\}:= \frac{\sum_{i=1}^m x_i e^{\beta x_i}}{\sum_{i=1}^m e^{ \beta x_i}},
\end{array}
\vspace{-2pt}
\end{equation}
where $\beta> 0$ is an adjustable parameter determining an under-approximation of the true minimum and maximum \cite{gilpin2020smooth}.
\section{Problem Statement}
Consider a dynamical system given by:\vspace{-2pt}
\begin{equation}
\label{eq:dynamics}\vspace{-2pt}
\begin{array}{l}
q^+(t)=f(q(t),u(t)),\\
q(0)=q_0,
\end{array}
\vspace{-2pt}
\end{equation}
where $q^+(t)$ stands for $\dot q(t)$ in continuous time and for $q(t+1)$ in discrete time, $q(t) \in \mathcal{Q} \subseteq\mathbb{R}^n$ is the state of the system and $u(t) \in \mathcal{U} \subseteq\mathbb{R}^m$ is the control input at time $t$, $q_0 \in \mathcal{Q} $ is the initial state and $f: \mathcal{Q} \times  \mathcal{U} \rightarrow  \mathcal{Q}$ is a Lipschitz continuous function. 
We denote the system trajectory generated by applying control input $\mathbf{u}$ for a finite time $T$ starting from the initial state $q_0$ by $\mathbf{q}(q_0,\mathbf{u})$, where $\mathbf{u}$ is a function of time 
or a discrete ordered sequence. 
Consider a cost function $J(u(t),q(t))$ and assume a desired temporal specification is given by a STL formula $\varphi$ over the system's trajectories. The control synthesis problem is formulated as:
\begin{problem}\label{problem formulation}
 Find an optimal control policy $\mathbf{u}^*$ that minimizes the cost function, and its corresponding system trajectory $\mathbf{q}(q_0,\mathbf{u}^*)$ satisfies $\varphi$ at time $0$:
\vspace{-3pt}
\begin{equation}
\label{eq:cost}
\begin{array}{c}
\mathbf{u}^*=\argmin _{u(t) \in \mathcal{U}} \; J(u(t),q(t)) \\
\text{s.t.   dynamics \eqref{eq:dynamics} are satisfied},\\
\mathbf{q}(q_0,\mathbf{u})\models^0 \varphi.
\end{array}
\end{equation}
\end{problem}
The authors of \cite{milp} showed that the optimization in (\ref{eq:cost}) can be mapped to a MILP if the cost and formula $\varphi$ are linear. In order to achieve robust satisfaction of $\varphi$ for systems with disturbances, 
by exploiting the \textit{soundness} property and considering the traditional robustness, 
later works re-formulated the control synthesis problem as \cite{raman,belta2019formal}:\vspace{-3pt}
\begin{equation}\small
\label{eq:opt-org}
\begin{array}{c}
\mathbf{u}^*={\argmax}_{u(t) \in \mathcal{U}} \; \rho(\varphi,\mathbf{q}(q_0,\mathbf{u}),0)-\lambda \; J(u(t),q(t))\\
\text{s.t.}\;\;\;
\text{dynamics} \;\; \eqref{eq:dynamics}\;\; \text{are satisfied}, \\
 {}\rho(\varphi,\mathbf{q}(q_0,\mathbf{u}),0)>0,
\end{array}
\vspace{-3pt}
\end{equation}
where $\lambda$ captures the trade-off between maximizing the robustness and minimizing the cost. The optimization problem \eqref{eq:opt-org} was solved using MILPs~\cite{raman} or gradient-based methods based on smooth approximations of $\rho$ which was applied to general nonlinear systems~\cite{pant2017smooth}. However, the traditional robustness only considered satisfaction of a formula at the most extreme sub-formula and time, hindering the optimization to find a more robust solution. Later works refined STL robustness by accumulating/averaging the robustness of all the sub-formulae over time \cite{discrete,iman19,varnai,gilpin2020smooth,ACC}.

In many applications, a high-level temporal logic specification may consist of 
obligatory or alternative sub-specifications or timings with different importance or priorities. The expressivity of traditional STL does not allow for specifying these preferences. 
Let $\varphi=\mathbf{F}_{[0,5]} (S > 0)$, which is satisfied if $S$ becomes greater than 0 within 5 time steps, and assume that satisfaction at earlier times within this deadline is more desirable. The traditional or average-based robustness have the same score for discrete-time signals $S_1=\{0,1,0,0,0,0\}$ and $S_2=\{0,0,0,0,0,1\}$, while it would be natural to assign a higher robustness to $S_1$ due to satisfaction of $\varphi$ at an earlier time. Imposing importance and priorities of satisfaction especially becomes important when a STL formula has conflicting obligatory specifications.

\begin{exm} \label{ex:car}
Consider a car driving on the two-lane road  
shown in Fig.~\ref{fig:ex1-conflict} \cite{rrt}. The car starts from an initial point at $t=0$ and has to reach \textit{Green} within 7 steps. Meanwhile, it has to always stay in its lane, and avoid the \textit{Blocked} area on the road. Assuming the duration of the overall task is bounded by 7, we formally define this specification as: $\varphi=\varphi_1 \land \varphi_2 \land \varphi_3$ where 
$\varphi_1= \mathbf{F}_{[0,7]}\textit{Green}$, $\varphi_2 = \mathbf{G}_{[0,7]} \neg \textit{Blocked}$, $\varphi_3=\mathbf{G}_{[0,7]} \textit{Lane}$.
As illustrated in Fig. \ref{fig:ex1-conflict}, in order to reach \textit{Green}, the car must either pass through the blocked area ($c_1\models \varphi_1$,$\varphi_3$ but $c_1 \nmodels \varphi_2$) 
or violate the lane requirement ($c_2\models \varphi_1,\varphi_2$, but $c_2 \nmodels \varphi_3$)
. In this example, a trajectory that can satisfy $\varphi$ does not exist. The minimally violating trajectory is dependent on the satisfaction importance of the obligatory tasks $\varphi_2$ and $\varphi_3$. 
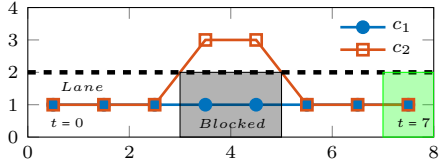
\begin{figure}[t]
\center
%
%
\definecolor{mycolor1}{rgb}{0.00000,0.44700,0.74100}%
\definecolor{mycolor2}{rgb}{0.85000,0.32500,0.09800}%
\definecolor{mycolor3}{rgb}{0.92900,0.69400,0.12500}%
\definecolor{mycolor4}{rgb}{0.49400,0.18400,0.55600}%
\definecolor{mycolor5}{rgb}{0.46600,0.67400,0.18800}%
\definecolor{mycolor6}{rgb}{0.30100,0.74500,0.93300}%
\pgfplotsset{tick label style={font=\scriptsize}}
\begin{tikzpicture}

\begin{axis}[%
width=2.1246028in,
height=0.6785754in,
at={(1.011in,0.642in)},
scale only axis,
xmin=0,
xmax=8,
ymin=0,
ymax=4,
axis background/.style={fill=white},
legend style={legend cell align=left, align=left, draw=none,fill=none}
]
\addplot [color=mycolor1, line width=1.0pt, mark=*, mark options={solid, fill=mycolor1, mark size=2.1pt}]
  table[row sep=crcr]{%
0.5	1\\
1.5	1\\
2.5	1\\
3.5	1\\
4.5	1\\
5.5	1\\
6.5 1\\
7.5 1\\
};
\addlegendentry{\scriptsize ${c}_1$}

\addplot [color=mycolor2, line width=1.0pt, mark=square, mark options={solid, fill=mycolor2, mark size=2.1pt}]
  table[row sep=crcr]{%
0.5	1\\
1.5	1\\
2.5	1\\
3.5	3\\
4.5	3\\
5.5	1\\
6.5 1\\
7.5 1\\
};
\addlegendentry{\scriptsize ${c}_2$}




\addplot [color=black,  dashed , line width=1.7]
  table[row sep=crcr]{%
  0 2\\
0.5	2\\
1.5	2\\
2.5	2\\
3.5	2\\
4.5	2\\
5.5	2\\
7 2\\
8 2\\
};

\addplot[area legend, draw=black, fill=black, fill opacity=0.3]
table[row sep=crcr] {%
x	y\\
3	0\\
5	0\\
5	2\\
3	2\\
}--cycle;
\addplot[area legend, draw=green, fill=green, fill opacity=0.3]
table[row sep=crcr] {%
x	y\\
7	0\\
8	0\\
8	2\\
7	2\\
}--cycle;

\node[right, align=left]
at (axis cs:3.173568482,0.407503) {\tiny $Blocked$};
\node[right, align=left]
at (axis cs:0.25345068482,0.437503) {\tiny $t=0$};
\node[right, align=left]
at (axis cs:7.0868482,0.4307503) {\tiny $t=7$};
\node[right, align=left]
at (axis cs:0.455068482,1.5627503) {\tiny $Lane$};

\end{axis}

\end{tikzpicture}%
\vspace{-4pt}
\caption{Trajectories $c_1$ and $c_2$ from Example \ref{ex:car}: the dots represent the positions at discrete times $t=0,1,\ldots,7$ (the continuous interpolation is shown for visualization).}
\label{fig:ex1-conflict}
\end{figure} 
\vspace{-3pt}
\end{exm}

In this paper, we extend the STL syntax and quantitative semantics to capture the importance and priorities of different sub-formulae and times. We show that the quantitative semantics of this extension is derived from the STL robustness. Hence, the optimization approaches described above, including MILPs and gradient-based methods, can be adapted to solve the synthesis problem for the proposed extended logic. 

\section{Weighted Signal Temporal Logic}
\label{sec:wstl}
In this section, we introduce 
wSTL
that enables the definition of user preferences (priorities
and importance).
\begin{definition}[wSTL Syntax]
\label{def:wstl}
The syntax of wSTL is an extension of the STL syntax, and is defined as:\vspace{-2pt}
\begin{equation}
\label{eq:wstl}
\varphi:= \top \mid \mu \mid \neg \varphi \mid {\bigwedge_{i=1:N}}^p \varphi_i \mid \mathbf{F}^\varpi_I \varphi \mid \mathbf{G}^\varpi_I \varphi,
\vspace{-2pt}
\end{equation}%
where the logical \emph{True} (and \emph{False})
value, the predicate $\mu$, and all the Boolean and temporal operators have the same interpretation as in STL.
The function $p: \{1, \ldots, N\} \to \mathbb{R}_{>0}$ assigns to each of the $N$ terms of the conjunction or disjunction (can be formed using conjunction and negation) a positive weight $p$; and
$\varpi: I \to \mathbb{R}_{>0}$ is a 
positive weight function for temporal properties. 
\end{definition}
The weights $p$ capture the {\em importance} of obligatory specifications or {\em priorities} of alternatives, respectively. The weights $\varpi$ capture satisfaction {\em importance} and {\em priorities} associated with \emph{always} and \emph{eventually} operators over the interval $I$, respectively. Higher values of $p$ and $\varpi$ correspond to higher importance and priorities. {\em Importance} will allow to weigh specifications that are all required to be satisfied (conjunctions for logical and always for temporal statements), while {\em priorities} will weigh specifications that accept alternative satisfactions (disjunctions for logical and eventually for temporal statements) (see Examples \ref{exp:importance}, \ref{exp:priority}, \ref{exp:pref}).
\begin{figure*}[bht!]
    \centering
    \subfloat[Importance of obligatory tasks\label{fig:wstl-examples-conjunction}]{%
      \scalebox{0.7}{
        \begin{tikzpicture}[scale=0.625, thick, >=stealth', dot/.style = {draw, fill = white, circle, inner sep = 0pt, minimum size = 4pt}]
          \coordinate (O) at (0,0);
          \draw[->] (-0.3,0) -- (8,0) coordinate[label = {below:$t$}] (xmax);
          \draw[->] (0,-0.3) -- (0,5) coordinate[label = {right:$S$}] (ymax);
          
          \fill[fill=gray!40] (2.0,3.0) rectangle (5.0,4.5);
          \draw[very thick] (2.0,3.0) {} -- (5.0,3.0) node[right] {$\mu_B=3.0$};
          \draw (5.0, 3.75) node[right] {$p_B = 2$};
          
          \fill[fill=gray!80] (1.0,0.0) rectangle (6.0,1.0);
          \draw[very thick] (1.0,1.0) node[left] {$\mu_A=1.0$} -- (6.0,1.0) {};
          \draw (6.0, 0.50) node[right] {$p_A = 4$};
          
          \draw[color=blue, domain=-0.0125:7]   plot (\x,{2 + 0.75 * sin(0.75 * \x r)}) {};
           \draw (1, 2.20) node[right,color=blue] {$s_s$};
           \draw[color=blue] (2.0, 2.75) node[dot, draw]{};
           \draw[color=blue] (6.0, 1.25) node[dot, draw]{};
           
            
            
            
            \draw[color=red, domain=-0.0125:7]   plot (\x,{2 + 1.30 * sin(0.8 * \x r)}) {};
            \draw (1, 3.60) node[right,color=red] {$s_v$};
             \draw[color=red] (2.0, 3.25) node[dot, draw]{};
            \draw[color=red] (6.0, 0.75) node[dot, draw]{};
            
          \draw[dashed] (1.0,-0.3) node[below]{$1$} -- (1.0,4.5) ;
          \draw[dashed] (2.0,-0.3) node[below]{$2$} -- (2.0,4.5) ;
          \draw[dashed] (5.0,-0.3) node[below]{$5$} -- (5.0,4.5) ;
          \draw[dashed] (6.0,-0.3) node[below]{$6$} -- (6.0,4.5) ;
          \draw[dashed] (-0.3,4.5) -- (3, 4.5) node[above] {$S_{max}=4.5$} -- (7.5,4.5) {};
      \end{tikzpicture}}}
    \hfill
  \subfloat[Priorities of alternatives\label{fig:wstl-examples-disjunction}]{%
      \scalebox{0.7}{
        \begin{tikzpicture}[scale=0.625, thick, >=stealth', dot/.style = {draw, fill = white, circle, inner sep = 0pt, minimum size = 4pt}]
          \coordinate (O) at (0,0);
          \draw[->] (-0.3,0) -- (8,0) coordinate[label = {below:$t$}] (xmax);
          \draw[->] (0,-0.3) -- (0,5) coordinate[label = {right:$S$}] (ymax);
          
          \fill[fill=green!20] (3.0,2.0) rectangle (6.0,4.5);
          \draw[very thick] (3.0,2.0) -- (6.0,2.0) node[right] {$\mu_B=2$};
          \draw (6.0, 3.5) node[right] {$p_B = 1$};
          
          \fill[fill=green!80] (4.0,0.0) rectangle (6.0,1.0);
          \draw[very thick] (4.0,1.0) -- (6.0,1.0) node[right] {$\mu_A=1.0$};
          \draw (6.0, 0.50) node[right] {$p_A = 10$};
          
          \draw[blue] (0.2, 2.5) node[]{} to[out=0,in=180] (5,0.5) node[dot] {};
           \draw (1, 1.80) node[right,color=blue] {$s_A$};
           
        \draw[red] (0.2, 2.5) node[]{} to[out=0,in=180] (5,1.3) node[dot] {};
        \draw[red] (3, 1.7) node[dot]{};
          \draw (1, 2.70) node[right,color=red] {$s'$};
          \draw (4.65, -.70) node[right] {$5$};
           
         \draw[blue] (0.2, 0.5) node[]{} to[out=0,in=180] (5,2.5) node[dot] {};
         \draw (1, 0.48350) node[right,color=blue] {$s_B$};
         
          
          \draw[dashed] (3.0,-0.3) node[below]{$3$} -- (3.0,4.5) ;
          \draw[dashed] (4.0,-0.3) node[below]{$4$} -- (4.0,4.5) ;
          \draw[dashed] (6.0,-0.3) node[below]{$6$}-- (6.0,4.5) {};
          \draw[dashed] (-0.3,4.5) -- (3, 4.5) node[above] {$S_{max}=4.5$} -- (7.5,4.5) {};
      \end{tikzpicture}}}
  \hfill  
  \subfloat[Preferences over time\label{fig:wstl-examples-weight-periodic}]{%
      \scalebox{0.7}{
        \begin{tikzpicture}[scale=0.625, thick, >=stealth', dot/.style = {draw, fill = white, circle, inner sep = 0pt, minimum size = 4pt}]
          \coordinate (O) at (0,0);
          \draw[->] (-0.3,0) -- (8,0) coordinate[label = {below:$t$}] (xmax);
          \draw[->] (0,-0.3) -- (0,5) coordinate[label = {right:$\varpi$}] (ymax);
          
          \draw[color=black, domain=1.0:6.0, samples=400]   plot (\x,{0.2 + 4.26 * (exp(- (\x -2) * (\x -2)/0.04) + exp(- (\x -3) * (\x -3)/0.04) + exp(- (\x -4) * (\x -4)/0.04) + exp(- (\x -5) * (\x -5)/0.04))}) {};
          \foreach \x/\xtext in {2, 3, 4, 5} 
            \draw (\x, 2pt) -- (\x, -2pt) node[anchor=north] {$\xtext$};
          \foreach \x/\xtext in {2/t_1, 3/t_2, 4/t_3, 5/t_4} 
            \draw[dashed] (\x, 0) -- (\x, 4.5) node[above] {$\xtext$};
          
          \draw[dashed] (1.0,-0.3) node[left]{$a=1$} -- (1.0,4.5) ;
          \draw[dashed] (6.0,-0.3) node[right]{$b=6$}-- (6.0,4.5) {};
          \draw[dashed] (-0.3,4.5) -- (7, 4.5) node[above] {$\varpi_{max}=4.5$} -- (7.5,4.5) {};
          \draw[dashed] (-0.3,0.2) -- (7, 0.2) node[above] {$\varpi_{min}=0.2$} -- (7.5,0.2) {};
      \end{tikzpicture}}}
    \hfill
  \subfloat[Preferences over time\label{fig:wstl-examples-weight-start-end}]{%
      \scalebox{0.7}{
        \begin{tikzpicture}[scale=0.625, thick, >=stealth', dot/.style = {draw, fill = white, circle, inner sep = 0pt, minimum size = 4pt}]
          \coordinate (O) at (0,0);
          \draw[->] (-0.3,0) -- (8,0) coordinate[label = {below:$t$}] (xmax);
          \draw[->] (0,-0.3) -- (0,5) coordinate[label = {right:$\varpi$}] (ymax);
          
          \draw[color=black, domain=1.0:6.0]  (1.0,4.5) to[out=0,in=180] (3.0,0.2) to[out=0,in=180] (4.5,0.2) to[out=0,in=180] (6.0,3.0);
  
          \draw[dashed] (1.0,-0.3) node[left]{$a=1$} -- (1.0,4.5) ;
          \draw[dashed] (6.0,-0.3) node[right]{$b=6$}-- (6.0,4.5) {};
          \draw[dashed] (-0.3,4.5) -- (7, 4.5) node[above] {$\varpi_{max}=4.5$} -- (7.5,4.5) {};
          \draw[dashed] (-0.3,3.0) -- (7, 3.0) node[above] {$\varpi_{end}=3.0$} -- (7.5,3.0) {};
          \draw[dashed] (-0.3,0.2) -- (7, 0.2) node[above] {$\varpi_{min}=0.2$} -- (7.5,0.2) {};
      \end{tikzpicture}}}
  \caption{wSTL performance for different specifications}
  \label{fig:wstl-examples}
\end{figure*}
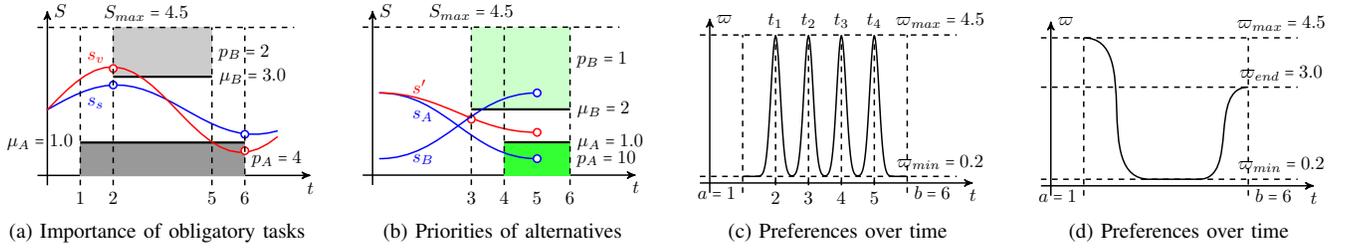
Throughout the paper, if the weight function associated with an operator (Boolean or temporal) in a wSTL formula is constant 1, we drop it from the notation.
Thus, STL formulae are wSTL formulae with all weights equal to 1.

The Boolean (qualitative) semantics of a wSTL formula is the same as the associated STL formula without the weight functions, i.e., $S \models^t \varphi \Leftrightarrow S \models^t \hat{\varphi}$,  
where $\hat{\varphi}$ is the unweighted version of a wSTL formula $\varphi$. 
\begin{definition}[wSTL Robustness]
\label{def:extended-robustness}
Given a wSTL specification $\varphi$ and a signal $S$,
the weighted robustness score $r^w(\varphi, S, t)$ at time $t$
is recursively defined as:
\begin{equation}
\label{eq:general-weighted-robustness}
\begin{aligned}
&r^w(\mu ,S,t) :=l(S(t)),\\
&r^w(\neg \varphi, S, t) : = - r^w(\varphi, S, t),\\
& r^w\left({\bigwedge_i}^p \varphi_i, S, t\right) := \otimes^\land(p, [r^w(\varphi_1, S, t), \ldots, r^w(\varphi_N, S, t)]),\\
& r^w\left({\bigvee_i}^p \varphi_i, S, t\right) := \oplus^\lor(p, [r^w(\varphi_1, S, t), \ldots, r^w(\varphi_N, S, t)]),\\
& r^w\left(\mathbf{G}^\varpi_I\varphi, S, t\right) := \mathop{\otimes^\mathbf{G}}(\varpi, r^w(\varphi, S, t + \cdot), I),\\
& r^w\left(\mathbf{F}^\varpi_I\varphi, S, t\right) := \mathop{\oplus^\mathbf{F}}(\varpi, r^w(\varphi, S, t + \cdot), I),
\end{aligned}
\vspace{-2pt}
\end{equation}
where $\otimes^\land$, $\oplus^\lor$, $\otimes^\mathbf{G}$, and $\oplus^\mathbf{F}$ are aggregation functions associated with the $\land$, $\lor$, $\mathbf{G}$ and $\mathbf{F}$ operators, respectively, which must satisfy
$\min\{x\} \cdot \otimes^\land(p, x) > 0$ and 
$\max\{x\} \cdot \oplus^\lor(p, x) > 0$ for all $x \in \mathbb{R}^N$, $x\neq 0$, and $p \in \mathbb{R}^N_{>0}$;
and
$\inf_{t\in I} R(t) \cdot \otimes^\mathbf{G}(\varpi, R) > 0$
and 
$\sup_{t\in I} R(t) \cdot \oplus^\mathbf{F}(\varpi, R) > 0$
for all $R : I \to \mathbb{R}_{\neq 0}$ and $\varpi: I \to \mathbb{R}_{>0}$.
\end{definition}


\begin{theorem}[wSTL Soundness]
\label{thm:wstl-soundness}
The weighted robustness score $r^w$ given by Def.~\ref{def:extended-robustness} is sound:\vspace{-2pt}
\begin{equation}
\begin{array}{l}
   r^w(\varphi, S, t)>0 \Leftrightarrow \rho(\hat\varphi, S, t)>0 \rightarrow S \models^t \varphi,\\
   r^w(\varphi, S, t)<0 \Leftrightarrow \rho(\hat\varphi, S, t)<0 \rightarrow S \nmodels^t \varphi
   \end{array}
   \vspace{-2pt}
\end{equation}
\end{theorem}\vspace{3pt}
\textbf{proof:}[Sketch]
A formal proof is omitted due to space constraints.
Informally, soundness can be viewed as a sign consistency between the weighted robustness $r^w$ and the (unweighted) traditional robustness $\rho$. The proof follows by structural induction and holds trivially for the base case corresponding to predicate formulae. The induction step also follows easily from the induction hypothesis and the constraints
placed on the aggregation functions in Def. \ref{def:extended-robustness}. Thus, the sign of the aggregation result correctly
captures the satisfaction and violation of
composite formulae connected via Boolean and temporal operators.

 The wSTL robustness must be defined such that $|r^w|$ is a measure of how much a wSTL specification is satisfied or violated, considering the importance or priorities of its sub-formulae and time. In the following, we define weighted generalizations of the traditional \cite{donze} and AGM \cite{ACC} robustness. The wSTL robustness for other compatible recursive STL robustness measures~\cite{pant2017smooth,discrete,iman19,varnai,gilpin2020smooth} can be defined similarly (See Sec.~\ref{case}). 
\subsection{Weighted Traditional Robustness}
The aggregation functions in a weighted generalization of the traditional robustness in~\eqref{eq:trad-robustness} 
can be defined as follows:
\begin{equation} \label{eq:wstl-trad}
 \begin{aligned}
& \otimes^\land(p, x) =  \min_{i=1:N}\{\big((\frac{1}{2}-{\bar{p}_i }) \sign(x_i) + \frac{1}{2}\big)\cdot x_i\}\\
&\oplus^\lor(p, x) = - \otimes^\land(p, -x)\\
&\otimes^\mathbf{G}(\varpi, R, I) = \inf_{t\in I}\{\big((\frac{1}{2}- {\bar{\varpi}(t) })\sign(R(t)) + \frac{1}{2}\big)\cdot R(t)\}\\
&\oplus^\mathbf{F}(\varpi, R, I) = - \otimes^\mathbf{G}(\varpi, -R, I)
 \end{aligned}
 \vspace{-3pt}
 \end{equation}
 where definitions for $\lor$ and $\mathbf{F}$ follow DeMorgan's law. $\bar{p}_i = \frac{p_i}{\sum_{j=1}^n p_j}$ and $\bar{\varpi}(t) = \frac{\varpi(t)}{\int_I \varpi(\tau)\,\mathrm{d}\tau}$ are normalized weights for the Boolean and temporal operators, respectively. $(1-\bar{p}_i)$ is interpreted as the total importance of all other subformulae except $\varphi_i$.
Therefore, in the case of satisfaction for conjunction,
${(1-\bar{p}_i)}r_i$ means that $r_i$ with $\bar{p}_i$ must be more important than
the hold-out importance of all other subformulae.
In the case of violation for conjunction, $\bar{p}_i r_i$ suggests
that violation of $\varphi_i$ has an importance of $\bar{p}_i$.
The same interpretation is given to $\bar{\varpi}(t)$ along the time interval $I$.

In the following, we discuss some examples to illustrate the expressivity of wSTL and weighted robustness. For brevity, we denote $r^w(\varphi, S, 0)$ by $r^w(\varphi, S)$.

\begin{exm}[Importance of obligatory tasks]\label{exp:importance}
Consider the wSTL specification $\varphi = \varphi_A \land^p \varphi_B = \mathbf{G}_{[1, 6]} (S \geq 1) \land^p \mathbf{G}_{[2, 5]} (S \leq 3)$
with $p_A=4$ and $p_B = 2$.
In Fig.~\ref{fig:wstl-examples-conjunction}, $s_s$ satisfies $\varphi$ and $s_v$ violates it.
The weights associated with the sub-formulae of the conjunction operator specify how important the satisfaction of
each obligatory task is, i.e., it is twice as important to stay above $1$ between time $t=1$ to $t=6$ than
to stay below $3$ from time $t=2$ to $t=5$.
If $r^w$ is defined as the weighted traditional robustness, we have $r^w(\varphi, s_s) = \min{(\frac{1}{3} \times 0.25}, \frac{2}{3}  \times 0.25)=0.083$ and $r^w(\varphi, s_v) = \min({\frac{2}{3}  \times -0.25}, \frac{1}{3}  \times -0.25)=-0.166$, highlighting the importance of $\varphi_A$ (it is more important for $s_s$ to satisfy $\varphi_A$, and violation of  $\varphi_A$ by $s_v$ is considered worse). 
\end{exm}
\begin{exm}[Priorities of alternative tasks]\label{exp:priority}
Consider the wSTL specification $\varphi = \varphi_A \lor^p \varphi_B = \mathbf{F}_{[4, 6]} (S\leq 1) \lor^p \mathbf{F}_{[3, 6]} (S\geq 2)$
with $p_A = 10$ and $p_B = 1$ and signals in Fig.~\ref{fig:wstl-examples-disjunction}. 
If $r^w$ is defined as the weighted traditional robustness, we have 
$r^w(\varphi, s_A) = \max({\frac{10}{11} \times 0.5}, \frac{10}{11} \times -0.8)=0.45$, while $r^w(\varphi, s_B) = \max(\frac{1}{11} \times -1.3, {\frac{1}{11} \times 0.5})=0.045$. Therefore, although both signals satisfy $\varphi$, $s_A$ is preferred to $s_B$, i.e., has a higher robustness, because it visits the higher priority region (defined by $\varphi_A$ within $[4, 6]$) while $s_B$ visits the lower priority region (induced by $\varphi_B$ within $[3, 6]$). Similarly, in the case of violation, weighted traditional robustness for the signal $s'$ is determined by the time $t=5$ (rather than $t=3$ which has the same distance from the lower priority region) since it is closer to (satisfy) the higher priority region. This leads to moving the signal towards satisfying $\varphi_A$ when maximizing robustness in the synthesis problem.
\end{exm}
\begin{exm}[Preferences over time]\label{exp:pref}
Consider the formulae $\varphi_F = \mathbf{F}^\varpi_{[1,6]} \varphi$ and $\varphi_G = \mathbf{G}^\varpi_{[1,6]} \varphi$.
Fig.~\ref{fig:wstl-examples-weight-periodic} and~\ref{fig:wstl-examples-weight-start-end} show two example weight functions.
For \emph{eventually}, $\varphi_F$ with weight $\varpi$ from Fig.~\ref{fig:wstl-examples-weight-periodic} specifies that the
task $\varphi$ should be done within $[1, 6]$ with higher priorities at one of the times $\{t_1, t_2, t_3, t_4\}$;
while the weight $\varpi$ in Fig.~\ref{fig:wstl-examples-weight-start-end} gives priorities to satisfaction at the endpoints especially at the start.
For \emph{always}, $\varphi_G$ with weight $\varpi$ from Fig.~\ref{fig:wstl-examples-weight-periodic} specifies that
$\varphi$ must hold at all times within $[1, 6]$, more importantly at times $\{t_1, t_2, t_3, t_4\}$;
while the $\varpi$ in Fig.~\ref{fig:wstl-examples-weight-start-end} specifies a higher importance at the end of the
interval and the highest importance at the start.
\end{exm}
\begin{figure*}[!thb]
\small
\begin{equation}\label{eq:agm-weighted-new}
\begin{split}
\eta^w\left({\bigwedge_i}^p \varphi_i, S, t\right) &:=\otimes^\land(p, [\eta^w(\varphi_1, S, t), \ldots, \eta^w(\varphi_N, S, t)])= \begin{cases}
\exp \left( {\sum\limits_{i } {\bar{p}_i} \ln ({{ \eta^w ({\varphi _i},S,t)} }})\right)  & \text{if } \forall i :\eta^w(\varphi_i,S,t) > 0, \\
{\sum\limits_{i } {{\bar{p}_i}{[\eta^w(\varphi_i, S, t)]_- }} }  &  \text{otherwise}
\end{cases} \\%
\eta^w \left( {{\mathbf{G}}^\varpi_{I}}\varphi, S, t \right) &:= \begin{cases}
\exp \left( \sum\limits_{t'\in t+I} \bar{\varpi}(t'-t) \ln ( \eta^w(\varphi, S, t')) \right)  & \text{if } \forall t' \in t+I : \eta^w ({\varphi},S,t') > 0,  \\
\sum\limits_{t'\in t+I} \bar{\varpi}(t'-t) [\eta^w(\varphi, S, t')]_- & \text{otherwise}
\end{cases}
\end{split}
\end{equation}
\end{figure*}
\subsection{Weighted AGM Robustness}
Consider a discrete-time system with time domain given by an ordered sequence $\tau:=\{k \mid k\in\mathbb{Z}_{\geq 0}\}$. We adapt the \emph{Arithmetic-Geometric Mean} (AGM) robustness to a wSTL weighted AGM robustness.  
The weighted AGM robustness captures the satisfaction of all sub-formulae and time, as well as their importance and priorities. For example, for $\varphi=\mathbf{F}^\varpi_{[0,5]} (S > 0)$, if satisfaction at earlier times is preferred, the weighted AGM robustness for $S_1=\{0,1,0,0,0,0\}$ is higher than $S_2=\{0,0,0,0,0,1\}$, but lower than $S_3=\{0,1,1,1,0,0\}$ since $S_3$ satisfies $\varphi$ 
as early as $S_1$ but also at more time points. Notice that weighted traditional robustness cannot distinguish between $S_1$ and $S_3$. Aggregation functions in weighted AGM robustness denoted by $\eta^w$ can be defined using weighted arithmetic- and geometric- means from Sec.~\ref{sec:wmean}. 
We define $\eta^w$ for conjunction and always operators recursively in \eqref{eq:agm-weighted-new}. 
Weighted AGM $\eta^w$ for other operators ($\oplus^\lor$ and $\oplus^\mathbf{F}$) can be defined accordingly by DeMorgan's law.

\begin{exm}
We demonstrate how the conjunction function changes for different normalized vectors $p$ for $\eta^w(\varphi_1  \land ^p \top , {S})$, where $\eta^w(\top , {S})=1$ is fixed, and $\eta^w(\varphi_1 , {S}) \in [-1,1]$. As illustrated in Fig. 3, by assigning a higher importance to $\top$
, $\eta^w(\varphi_1  \land^p \top , {S})$ is closer to $1$, and for a higher importance to $\varphi_1$, robustness is closer to $\eta^w(\varphi_1, {S})$. Similar to the AGM robustness, the weighted AGM robustness $\eta^w(\varphi,S,t)$ 
is sound and monotone, and 
for $\eta^w(\varphi_1 , {S})=1$, we have $\eta^w(\varphi_1  \land^p \top , {S}) =1$ independent of $p$.
\begin{figure}
\label{fig:conjunction}
\centering
\input{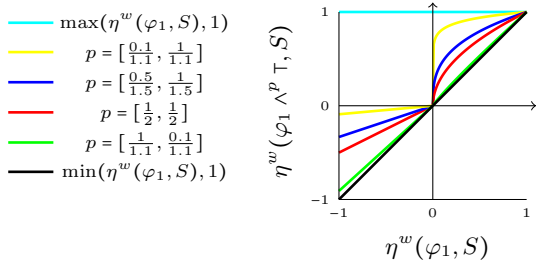}
\vspace{-4pt}
\caption{Effect of $p$ in $\eta^w(\varphi_1  \land^p \top , {S})$
, legends correspond to the signals from top to bottom.} 
\end{figure} 
\end{exm}
\begin{exm}\label{ex2}
Consider the discrete-time signals $S_4,S_5,S_6$ shown in Fig. 
\ref{fig:ex2} and $\varphi={\mathbf{F}^\varpi}_{[0,3]} (S \ge 0)$
. We can choose the weights (priorities) as $\varpi(t)= \gamma ^ {(t-1)}$ with the discount factor $\gamma$ to reward the satisfaction of the formula at earlier time steps within the deadline. For larger $\gamma$ (closer to $1$), satisfaction at different time points is considered to have similar priorities, and by decreasing $\gamma$, satisfaction at earlier times within the deadline results in a higher weighted AGM robustness, as seen in Table \ref{table_F}. Note that the unweighted robustness definitions cannot distinguish these signals.
\begin{figure}[!bht]
\includegraphics[width=8.5cm]{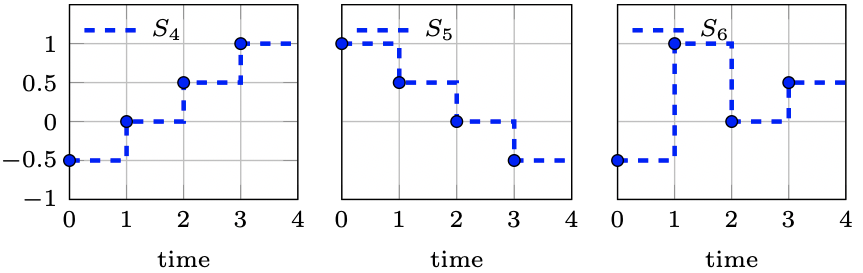}
\vspace{-3pt}
\caption{Discrete-time signals in Example 
\ref{ex2}.}
\label{fig:ex2}
\end{figure}
\begin{table}[!tbh]
\caption{Traditional and weighted AGM robustness for different values of $\gamma$ in $\varpi(t)= \gamma ^ {(t-1)}$ for $\varphi={\mathbf{F}}^\varpi_{[0,3]} (S \ge 0)$}
\label{table_F}
\centering
\begin{tabular}{ccccccc}
\hline 
Signal & $\rho$ &  $\eta$ & $\eta^w, \gamma=0.9$ & $\eta^w, \gamma=0.5$ & $\eta^w, \gamma=0.1$\\
\hline
\addlinespace
$S_4$ & 1  & 0.375 & 0.330 &   0.133 & 0.005\\

$S_5$ & 1  & 0.375 & 0.420 &  0.666 & 0.945\\

$S_6$ & 1  & 0.375 & 0.367 & 0.300 & 0.090\\
\hline
\end{tabular}
\end{table}

\vspace{-3pt}
\end{exm}
\begin{exm}
Consider Example \ref{ex:car} with trajectories $c_1$ and $c_2$. Assuming $\varpi=\mathbf{1}$, from \eqref{eq:agm-weighted-new} we have $\eta^w(\varphi_2,c_1)=\eta^w(\varphi_3,c_2)= -\frac{2}{8}$. Similarly, for the overall specification $\varphi$, we have $\eta^w(\varphi, c_1) =\sum_{i = 1}^3 {\bar{p}_i{[\eta^w(\varphi_i, c_1)]_-  }} ={\bar{p}_2\; \eta^w(\varphi_2,c_1)}$ 
and $\eta^w(\varphi, c_2)=\bar{p}_3 \; \eta^w(\varphi_3,c_2)$. If $p_2=p_3$, satisfaction of both $\varphi_1$ and $\varphi_2$ will have the same importance. Thus, $c_1$ and $c_2$ have the same robustness. Assume avoiding \textit{Blocked} is more important than staying in the lane. By choosing $p_2 > p_3$, we can emphasize the importance of $\varphi_2$. As a result, $\bar{p}_2 > \bar{p}_3$ and $\eta^w(\varphi, c_1)<\eta^w(\varphi, c_2) < 0 $. Since the satisfaction of all the sub-formulae is not feasible, 
$c_2$ is the minimally violating trajectory (compared to $c_1$). \vspace{-3pt}
\end{exm}
\section{Synthesis Using Weighted Robustness}
The \emph{soundness} property of the weighted robustness $r^w$ allows us to reformulate the synthesis problem \eqref{eq:opt-org} as:\vspace{-2pt}
\begin{equation}\small
\label{eq:opt}
\begin{array}{c}
\mathbf{u}^*={\argmax}_{u(t) \in \mathcal{U}} \; r^w(\varphi,\mathbf{q}(q_0,\mathbf{u}))-\lambda \; J(u(t),q(t))\\
\text{s.t.}\;\;\;\;
\text{dynamics} \;\; \eqref{eq:dynamics}\;\; \text{are satisfied}, \\
 {}r^w(\varphi,\mathbf{q}(q_0,\mathbf{u}))>\epsilon,
\end{array}
\vspace{-2pt}
\end{equation}
where $\epsilon \geq 0$ is the lower bound of the satisfaction margin (soundness threshold) as captured by the weighted generalization of the recursive STL robustness \cite{pant2017smooth}. 
Since the weighted robustness of wSTL is defined as a generalization of an unweighted recursive STL robustness, previous robustness optimization frameworks can be adapted to solve \eqref{eq:opt}.

%

We consider a discrete-time system with a finite time $T$ and assume the control policy to be synthesized is a discrete ordered sequence $\mathbf{u}=\small{u(0)u(1)\cdots u(T-1)}$. We use a gradient-based optimization to solve \eqref{eq:opt}. Note that the weights do not increase the computation time compared to the unweighted robustness optimization. $r^w$ is iteratively maximized by updating the input variable $u(t)$ at each time $t$ proportional to the gradient of $r^w$ such that $\small{u(t)^{l+1}\gets u(t)^{l}+\alpha^l\; \nabla r^w }$, 
where $l$ is optimization iteration, $\alpha^l$ is step size and $\nabla r^w = \frac{\partial r^w(\varphi,\mathbf{q}(q_0,\mathbf{u}))}{\partial u(t)}$ 
\cite{GAbook}. Therefore, depending on the aggregation functions in $r^w$, the weights $p$ and $\varpi$ associated with $\varphi$ affect the gradient and optimization. A similar synthesis framework can be applied to continuous-time systems with a Zeroth-Order Hold (ZOH) input \cite{CDC19}. 
\section{Case Study}
\label{case}
Consider a discrete-time nonlinear dynamical system as:\vspace{-3pt}
\begin{equation}
\label{eq:nonlinear}
\begin{array}{l}
x(t+1)=x(t)+\cos\theta(t)v(t),\\
y(t+1)=y(t)+\sin\theta(t)v(t),\\
\theta(t+1)=\theta(t)+v(t)w(t),
\end{array}
\vspace{-3pt}
\end{equation}
and a task \enquote{\textit{Eventually} visit \textit{A} or \textit{B} within $[1,10]$ \textit{and} \textit{eventually} visit $\textit{C}$ within $[11,20]$ \textit{and} \textit{Always} avoid \textit{Unsafe} \textit{and} \textit{Always} stay inside \textit{Boundary}} given by wSTL formula:\vspace{-3pt}
\begin{equation}
\label{eq:ex3}
\begin{array}{l}
\varphi=
(\mathbf{F}_{[1,10]}\;(\textit{A}\vee^p \textit{B})) \;\wedge(\mathbf{F}_{[11,20]} \textit{ C}) \\ \;\;\;\;\;\;\;\;\;\;\;
\wedge\;(\mathbf{G}_{[1,20]}\;\neg\textit{Unsafe}) \;\wedge(\mathbf{G}_{[1,20]}\;\textit{Boundary}) ,
\end{array}
\vspace{-3pt}
\end{equation}
where $\textit{A}=[7,9] \times [1,3]$ or $\textit{B}=[1,3] \times [7,9]$ and $\textit{C}=[7,9]^2$ are regions to be sequentially visited within the associated deadlines, $\textit{Unsafe}=[3,6]^2$ and $\textit{Boundary}=[0,10]^2$. $q=[x,y,\theta]$ is state vector  
with initial state $q_0=[1,1,\pi/4]$, $u=[v,w]$ 
is the input vector with $\mathcal{U}=[-2,2]^2$, and cost function is
$J=\frac{1}{2} \sum_{t=0}^{T-1} {\|u(t)\|^2}$ with $T=20$, $\lambda=0.05$ in \eqref{eq:opt}.
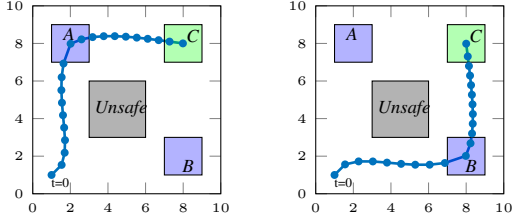
\begin{figure}[t]
\centering
\begin{tabular}{cc}
%

\definecolor{mycolor1}{rgb}{0.00000,0.44700,0.74100}%
\definecolor{mycolor2}{rgb}{0.00000,0.44706,0.74118}%
\definecolor{mycolor3}{rgb}{1.00000,1.00000,0.00000}%
\pgfplotsset{tick label style={font=\tiny}}
\begin{tikzpicture}

\begin{axis}[%
width=2.5cm,
height=2.5cm,
at={(0cm,0cm)},
scale only axis,
xmin=0,
xmax=10,
ymin=0,
ymax=10,
axis background/.style={fill=white},
]

\addplot [color=mycolor1, line width=1.0pt, mark=*, mark options={solid, fill=mycolor2, mark size=1pt}]
  table[row sep=crcr]{%
1	1\\
1.53605767415167	1.53605767415167\\
1.69799508607259	2.18358606239048\\
1.71524231842689	2.85188529791927\\
1.67000026557609	3.51937574003804\\
1.60784581991524	4.1813532230405\\
1.55293511958309	4.84345884963946\\
1.5224501074102	5.51269570889177\\
1.53714132799138	6.19952117018687\\
1.63636614618946	6.93084463944861\\
2.01293050499245	7.98355263411528\\
2.59132991914244	8.21628724629761\\
3.18191812773119	8.33633203474031\\
3.77328168693125	8.38529996225051\\
4.36240396169636	8.38756862416967\\
4.94863434894032	8.35869164153382\\
5.53223986705875	8.30905024065948\\
6.11383947766407	8.24578155443991\\
6.69411058220688	8.17399907874335\\
7.2745439104093	8.09749967526495\\
7.98557419641288	8.00213194975697\\
};

\addplot[area legend, draw=black, fill=blue, fill opacity=0.3]
table[row sep=crcr] {%
x	y\\
3	9\\
1	9\\
1	7\\
3	7\\
}--cycle;

\addplot[area legend, draw=black, fill=blue, fill opacity=0.3]
table[row sep=crcr] {%
x	y\\
9	3\\
7	3\\
7	1\\
9	1\\
}--cycle;

\addplot[area legend, draw=black, fill=green, fill opacity=0.3]
table[row sep=crcr] {%
x	y\\
9	9\\
7	9\\
7	7\\
9	7\\
}--cycle;

\addplot[area legend, draw=black, fill=black, fill opacity=0.3]
table[row sep=crcr] {%
x	y\\
6	3\\
3	3\\
3	6\\
6	6\\
}--cycle;

\node[right, align=left]
at (axis cs:.5,.6) {\tiny t=0 };

\node[right, align=left]
at (axis cs:7.4608482,1.4687503) {\scriptsize \textit{B}};
\node[right, align=left]
at (axis cs:1.13585068482,8.48757503) {\scriptsize \textit{A}};
\node[right, align=left]
at (axis cs:7.6868482,8.414307503) {\scriptsize \textit{C}};
\node[right, align=left]
at (axis cs:2.80935068482,4.527503) {\scriptsize \textit{Unsafe}};
\end{axis}
\end{tikzpicture}%
&
%

\definecolor{mycolor1}{rgb}{0.00000,0.44700,0.74100}%
\definecolor{mycolor2}{rgb}{0.00000,0.44706,0.74118}%
\definecolor{mycolor3}{rgb}{1.00000,1.00000,0.00000}%
\pgfplotsset{tick label style={font=\tiny}}
\begin{tikzpicture}

\begin{axis}[%
width=2.5cm,
height=2.5cm,
at={(0cm,0cm)},
scale only axis,
xmin=0,
xmax=10,
ymin=0,
ymax=10,
axis background/.style={fill=white},
]

\addplot [color=mycolor1, line width=1.0pt, mark=*, mark options={solid, fill=mycolor2, mark size=1pt}]
  table[row sep=crcr]{%
1	1\\
1.56144238863363	1.56144238863363\\
2.2807898039243	1.72167139485058\\
3.03611479403101	1.71901527136023\\
3.78949782507287	1.65743506106726\\
4.53895977021117	1.59055735409251\\
5.29200211386475	1.54528973344951\\
6.05883165466201	1.54605643917104\\
6.86570661613353	1.63629340930963\\
7.98261857767241	2.01449431238479\\
8.2316638561919	2.68499640256367\\
8.3123966070569	3.20640179021866\\
8.34938155727992	3.72580909567778\\
8.35507021593691	4.24271127340857\\
8.33810485808499	4.75715853590466\\
8.30473239355465	5.26944701815127\\
8.25959543173924	5.77999347767865\\
8.20624848391728	6.28924820930512\\
8.14752156677595	6.79766527730444\\
8.08552866581754	7.30782542724239\\
8.00190576862338	7.98617953413718\\
};

\addplot[area legend, draw=black, fill=blue, fill opacity=0.3]
table[row sep=crcr] {%
x	y\\
3	9\\
1	9\\
1	7\\
3	7\\
}--cycle;

\addplot[area legend, draw=black, fill=blue, fill opacity=0.3]
table[row sep=crcr] {%
x	y\\
9	3\\
7	3\\
7	1\\
9	1\\
}--cycle;

\addplot[area legend, draw=black, fill=green, fill opacity=0.3]
table[row sep=crcr] {%
x	y\\
9	9\\
7	9\\
7	7\\
9	7\\
}--cycle;

\addplot[area legend, draw=black, fill=black, fill opacity=0.3]
table[row sep=crcr] {%
x	y\\
6	3\\
3	3\\
3	6\\
6	6\\
}--cycle;

\node[right, align=left]
at (axis cs:.5,.6) {\tiny t=0 };

\node[right, align=left]
at (axis cs:7.4608482,1.4687503) {\scriptsize \textit{B}};
\node[right, align=left]
at (axis cs:1.13585068482,8.48757503) {\scriptsize \textit{A}};
\node[right, align=left]
at (axis cs:7.6868482,8.414307503) {\scriptsize \textit{C}};
\node[right, align=left]
at (axis cs:2.80935068482,4.527503) {\scriptsize \textit{Unsafe}};

\end{axis}
\end{tikzpicture}%
 \end{tabular}
 \vspace{-5pt}
 \caption{Trajectories from the synthesized control $\mathbf{u}^*$ satisfy $\varphi$ and minimize the cost. $p_{\textit{A}}>p_{\textit{B}}$ (left), $p_{\textit{A}}<p_{\textit{B}}$ (right).}
\label{fig:syn} 
\end{figure} 

By defining $\otimes^\land(p, x) = \widetilde{\min}_{i=1:N}\{\big((\frac{1}{2}-\bar{p_i}) \sign(x_i) + \frac{1}{2}\big)\cdot x_i\}$,
$\oplus^\lor(p, x) =\widetilde{\max}_{i=1:N}\{\big(-(\frac{1}{2}-\bar{p_i}) \sign(x_i) + \frac{1}{2}\big)\cdot x_i\}$ 
from \eqref{eq:smooth} (similarly for $\oplus^\mathbf{G}$ and $\oplus^\mathbf{F}$), and approximating $\sign$ by $\sign(x)\simeq \tanh({\beta x})$, we obtain the weighted sound smooth robustness of \cite{gilpin2020smooth}, which is adjustable to a desired locality and masking level. 
Fig. \ref{fig:syn} shows trajectories obtained from optimizing \eqref{eq:opt} considering the weighted robustness of \cite{gilpin2020smooth} with $\beta=10$, achieved up to the same termination criteria with different priorities for visiting \textit{A} or \textit{B} as 
$p_{\textit{A}}=2$, $p_{\textit{B}}=1$ (left), and $p_{\textit{A}}=1$, $p_{\textit{B}}=2$ (right). The optimization is implemented in Matlab using the SQP optimizer and takes about $1.2$ seconds on a Mac with 2.5 GHz Core i7
CPU 16GB RAM. 
For the given symmetrical configuration and initial state, optimizing the weighted robustness ensures that the optimal trajectory visits the higher priority region \textit{A} or \textit{B} as chosen by the disjunction aggregator priorities $p$. 

\section{Conclusion And Future Work}
\label{sec:conclusion}

We presented an extension of STL to improve its expressivity by encoding the importance or priorities of sub-formulae and time in a formula. The new formalism, called wSTL, is advantageous especially in the problems where satisfaction of a formula is not feasible, and as a result, the less important sub-formulae or time are preferred to be violated to guarantee that more important ones are satisfied. The weighted robustness associated with wSTL also improved the optimal behavior in a control synthesis framework solved using gradient techniques where prioritized tasks were critical. Future work will investigate computationally efficient implementations of the wSTL robustness optimization, and learning frameworks for systematic design of weights to capture hierarchies of wSTL formulae consistently.  

\bibliographystyle{IEEEtran}
\bibliography{thebibliography}
\end{document}